\documentclass[letterpaper, 10 pt, conference]{ieeeconf}  

\IEEEoverridecommandlockouts                              

\usepackage{stfloats}
\usepackage[utf8]{inputenc}
\usepackage{graphicx} 
\usepackage{color}
\usepackage{comment}
\usepackage{amssymb} 
\usepackage{amsmath}
\usepackage{multirow}
\usepackage{float}
\usepackage{bm}
\usepackage{cancel}
\usepackage{subfigure}
\usepackage[dvipsnames]{xcolor}
\usepackage{cite}

\title{\LARGE \bf
A flexible propelled arm: Mechanical considerations for the use in UAVs}

\author{F. Ruiz, B. Arrue*, A. Ollero,~\IEEEmembership{GRVC Aerial Robotics Lab,~University of Seville}
\thanks{*Corresponding author: barrue@us.es
\newline
Other authors: frvincueria@us.es  aollero@us.es}
\thanks{Departamento de Ingeniería de Sistemas y Automática, Escuela Técnica Superior de Ingeniería, Camino de los Descubrimientos s/n, 41092 Sevilla}%
}

\begin{document}

\maketitle
\thispagestyle{empty}
\pagestyle{empty}

\begin{abstract}

This paper presents a soft propelled arm, 3D-printed in TPU 70A, designed to be used in flexible UAVs, an advantageous concept since it reduces risk in the event of a collision thanks to a high energy absorption. The proposed design also allows the possibility to adapt to the environment and land on pipelines. The  flexibility  of  the  arm  can  be  controlled  during the  additive  manufacturing  process  by  adjusting  the  infill rate or internal density. In this work, a mechanical design based on simulations with an experimentally adjusted 5-parameter Mooney-Rivlin non-linear model is proposed. Thrust efficiency is also maximized through CFD simulations.


\end{abstract}

\section{INTRODUCTION}

Unmanned Aerial Vehicles (UAVs) offer a cost-effective solution to several operations such as surveillance, monitoring, and inspection \cite{Bernard2011AutonomousTA,Ollero2012,OlleroManipulation}. The drone industry is expected to continue to grow exponentially in the coming years \cite{Rao2016} and therefore will have a huge impact on various sectors. The main reason is that this technology enables access to difficult-to-reach areas. In addition, they have the possibility of physically interacting with the environment, for example through robotic arms \cite{aerialmanipulator}.




However, their ability to interact with the environment, especially with humans, is limited \cite{Zheng2021EvolutionaryHC,ZHAO2020}. They are tipically made of rigid underlying structures. Soft robotics is dedicated to the design and construction of robots with physically flexible bodies \cite{Trivedi2008SoftRB, Rus2015DesignFA}. In the case of UAVs this trend can be applied to certain parts of the platform, reducing the risk that its structure can produce to human operators. It can also allow access to zones which are unreachable for humans and traditional UAVs, by modifying their own geometry during flight. This technique is known as morphing \cite{Riviere2018AgileRF} and can also be used to improve the aerodynamic properties and flight mechanics of the UAV \cite{Ajanic2020BioInspiredSW}. The most widespread concept for these changes in geometry is reversible plasticity, which can be achieved by temperature control \cite{ReversiblePlasticity}.

Recent soft aerial platforms developments include a micro quadcopter with flexible arms made of a layered structure, which gives enough rigidity to grant that the platform flies, while being sufficiently flexible so that in case of collisions, the arms fold and prevent them from breaking \cite{Floreano_origami}. Origami techniques have been also used to fold and unfold the wings of a fixed-wing aircraft \cite{iros-2016}.

\begin{figure}
  \begin{subfigure}[Lower surface]{
    \begin{minipage}[t]{0.99\linewidth}
    \centering
    \includegraphics[width=\textwidth,scale=1]{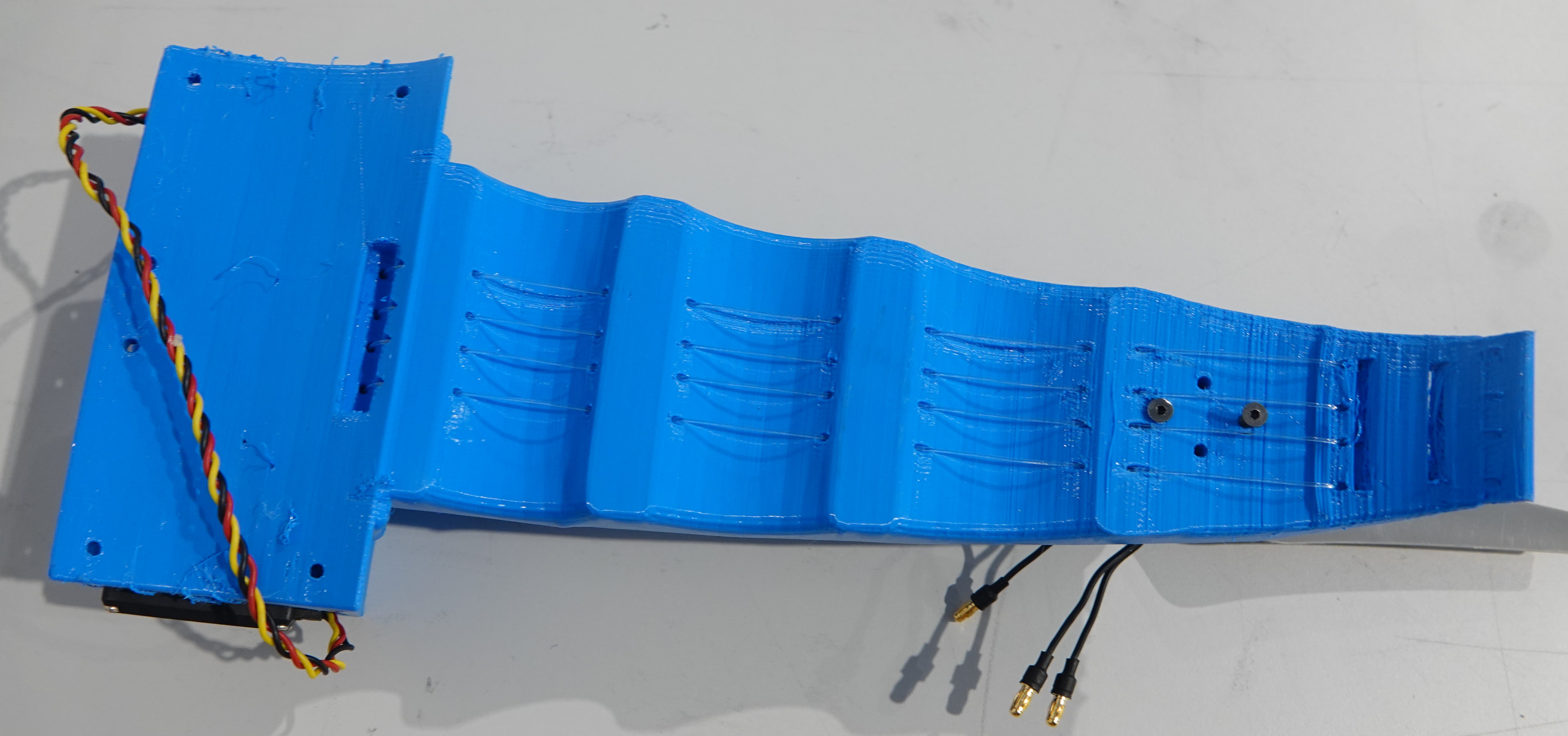}
    \end{minipage}%
    }%
  \end{subfigure}
  \begin{subfigure}[Upper surface]{
    \begin{minipage}[t]{0.99\linewidth}
    \centering
    \includegraphics[width=\textwidth,scale=1]{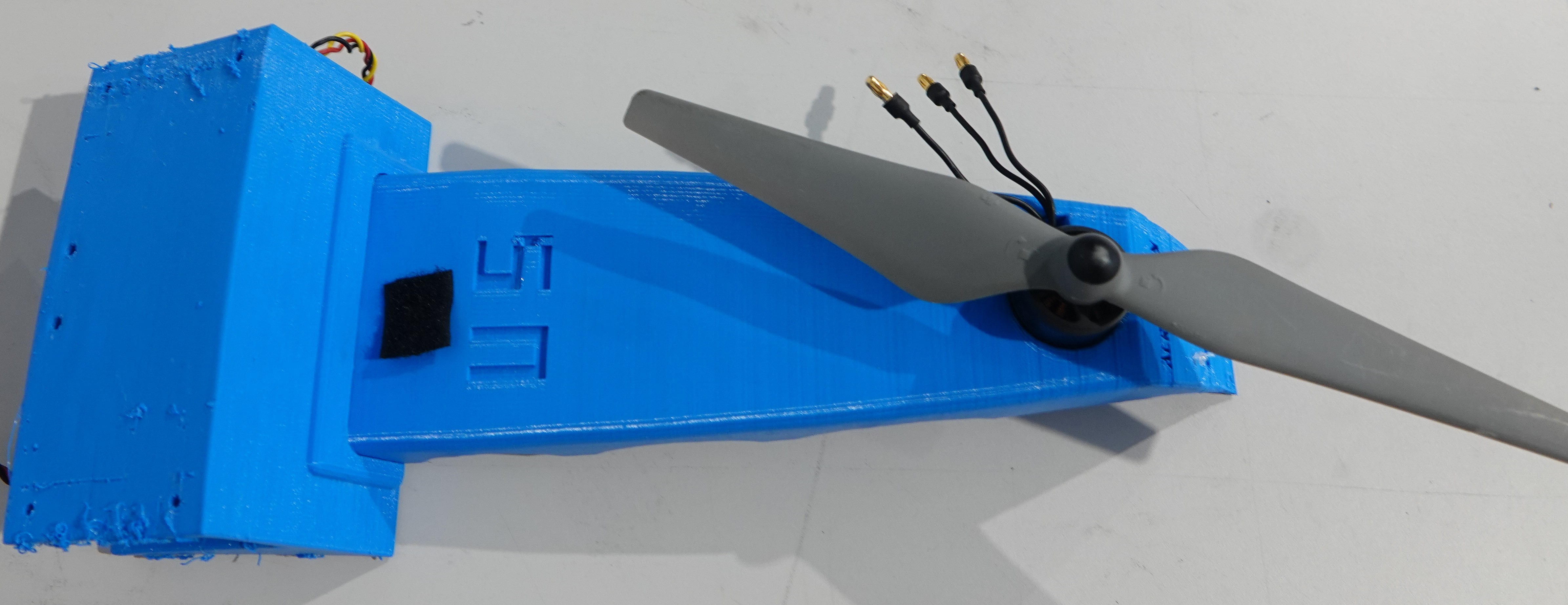}
    \end{minipage}%
    }%
  \end{subfigure}
 \caption{Soft propelled arm prototype 3D-printed in flexible TPU 70A}
 \label{f:UAV}
\end{figure}

In addition to a flexible structure, there are drones that also use soft actuators or sensors. For instance, robotic arms can employ soft-end effectors for contact manipulation \cite{Xiang2019SoftsmartRE}. In the field of soft actuators, TCP tensors  act like muscles and provide a greater force-weight ratio than traditional servos \cite{TammTCP}. In addition, soft systems can be used to perform perching on complex structures or pipelines, such as using tentacles \cite{GRubiales2021}. Even so, the term soft aerial robotics usually refers to unmanned vehicles that use soft materials in a generic way in their structure, giving them tremendous potential for interaction with humans. 

These soft components are tipically 3D-printed using novel materials such as Filaflex or Thermoplastic polyurethane (TPU), which are flexible and elastic. Other soft structures \cite{Mazzolai2012} are based on highly flexible silicones, like Ecoflex, and pneumatic actuators, which combined with specific internal structures, produce complex movements. In this work, the shore hardness scale of the ASTM D2240-00 standard is used to determine how flexible a material is.

The variability of the mechanical properties of this type of materials depend on the manufacturing techniques. Thus,  infill rate, nozzle temperature and layer height, have been studied by several authors \cite{TPUproperties,lee2017fundamentals}. In this work it has been decided to obtain experimentally-measured properties for each manufactured arm.

The modeling of rubber-like materials requires nonlinear elasticity theory. The study of the most appropriate models for TPU has been studied before \cite{hyperelastic2}, although the suitability of each model depends in turn on the application and the printing characteristics. In this work, a Mooney-Rivlin model is proposed. The behavior is determined by obtaining 5 parameters which can be obtained experimentally from the stress-strain curve \cite{hyperelastic}.

The design of a flexible propelled arm requires aerodynamic studies. The aerodynamic interaction between propellers and objects has been studied in deep \cite{CFD2}. The objective of this work is not to investigate CFD methodologies or models, but to study the specific aerodynamic effects that the present design experiences, in addition to evaluating the thrust losses due to the aerodynamic resistance of the arm.

The Multiple Reference Frame (MRF) method is a steady-state approximation suitable to study the fluid flow around a propeller with solid interaction \cite{CFDmethod}. In this method, the computational domain is split into two
regions: a rotating domain containing the propeller and a
stationary domain containing the flexible arm.

In this article, a flexible propelled arm 3D-printed in Thermoplastic Polyurethane (TPU 70A) is proposed. It has been designed with the aim of being integrated into a UAV prototype entirely 3D-printed in flexible filament, considerably reducing weight and increasing adaptability. The flexibility of the arm can be controlled during the additive manufacturing process by adjusting the internal density of the material $\rho_{TPU}$. 

The arm concept has been equipped with soft tendons to attach to the surface of elevated pipelines to be inspected. This future application has emerged as a continuation of  the  work  of  the  authors  within  the  framework  of  the European project HYFLIERS. The use of tentacles to attach to surfaces has been studied in previous works developed by the authors of this design \cite{GRubiales2021,PRamon2019}. However, in this work the concept is applied to a propelled arm, which leads to novel mechanical challenges.


The paper is structured as follows: Section II characterizes in depth the behavior of the flexible material and introduces the soft propelled arm concept. In Section III, the structural and aerodynamic analyses are described. Section IV shows the comparison between simulations and experimental results obtained in a test bench. Conclusions and new lines of research are presented in Section V.

\section{System overview}

\subsection{Soft propelled arm concept}

Finding a balance between the advantages provided by a flexible and deformable arm, and its mechanical robustness to be integrated into a UAV, is a considerable challenge. This work aims to find a mechanical design that minimizes the deflections of the arm in flight conditions, maintaining its ability to deform and adapt to pipes.

Therefore, the question is to what extent the effects of said flexibility do not affect the thrust efficiency of the arm. The goal is to achieve a minimum efficiency of 85 $\%$ in the whole operation range of the motors. At the same time, it is intended to minimize the energy needed by the servomotors to actuate the arm tendons.

Another relevant area of research is the additive manufacturing technique and its influence not only on the mechanical behavior, but also on the printing repeatability and the effects of long-term fatigue on soft bodies. Internal density, grid type or even the relative position between the structure and the extruder can significantly influence the behavior of the body by varying the direction in which the fibers are arranged. 

\begin{figure}
\includegraphics[width=0.47\textwidth,scale=0.25]{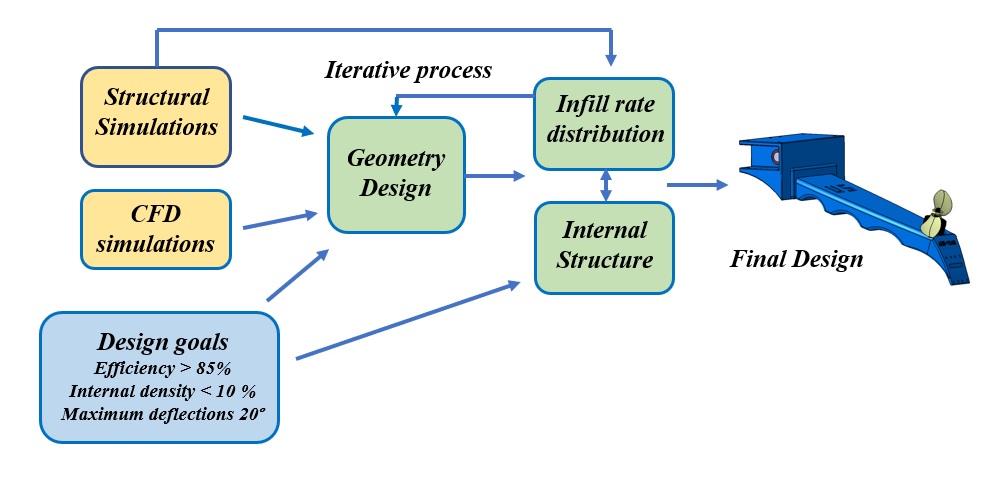}
\caption{Soft propelled arm design process }
\label{process}
\end{figure}


The soft arm (Figure \ref{f:UAV}) is designed to solve those problems. It is completely soft, excluding only the motorization and batteries. This is, by definition, a flexible material with a hardness of less than 70A on the shore hardness scale of the ASTM D2240-00 standard. For this purpose, the most suitable technique is 3D additive manufacturing. The design process followed in this work is explained in Figure 2.

\subsection{Soft material characterization}

Soft robotics is an area of knowledge closely related to materials science. The study of the mechanical behavior of this type of robots requires to analyze the internal structure of the material. In this work, different internal structures of the proposed material are analyzed numerically and experimentally.

Young’s modulus and the stress-strain curve of flexible filaments are provided by the manufacturer. However, as aforementioned, the real behavior of the structure varies depending on the parameters of additive manufacturing, especially the infill rate or internal density.

The corrected mechanical properties of the structure can be obtained experimentally through the stress-strain curves. First, the flexural elasticity modulus has been obtained through a simplified material strength model (cantilever beam model, Eq. 1). The tests are performed with the arm fixed at one end, while applying a series of forces at the other end, and measuring the displacements.

\begin{equation}
\label{desp_voladizo}
\delta = \frac{F L^3}{3 E I}
\end{equation}

In this equation, E is the desired elastic modulus, F is the applied force, L is the length of the arm and I is the inertia of the arm cross section. Therefore, from a set of experiments for different F-$\delta$ pairs, the slope $\alpha$ can be extracted and the elastic modulus is obtained as:

\begin{equation}
\label{Young}
E = \frac{\alpha L^3}{3 I}
\end{equation}

\begin{figure}
\includegraphics[width=0.47\textwidth,scale=0.25]{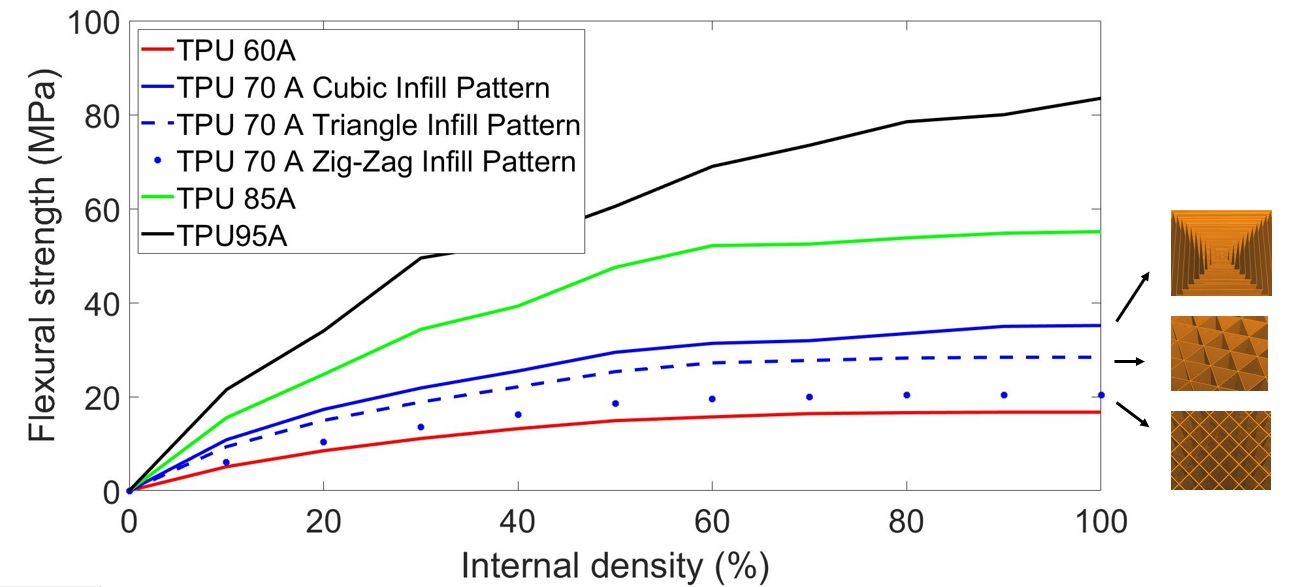}
\caption{Flexural modulus evolution as a function of infill rate for different TPU and internal grid types }
\label{flexural}
\end{figure}

Note that these values are only an initial approximation since the curve is nonlinear, and a linear interpolation is performed. Experimental results for various types of TPU are shown in Figure \ref{flexural}. In addition, the effect of the internal structure is also analyzed. The cubic arrangement of the filament layers provides the highest elasticity modulus. In contrast, the zig-zag case provides the largest deflections for the same force. These results should be considered in later design stages.

The selected filament is TPU 70 A, which is flexible while having good mechanical properties such as efficient vibration dampening and good impact resistance. Meanwhile, it is difficult to print, with the risk of blobs and stringing formation. Another disadvantage is the difficulty of maintaining a high repeatability between prints of the same design, affecting the elastic behavior.

On the other hand, the aforementioned F-$\delta$ curves can be directly related to strains and stresses. This is done through the equilibrium equations ($\sigma_{ii,i}+F_i=0$) and the compatibility equations between displacements and deformations ($\epsilon_{ii}=\delta_{i,i}$), which are applied to a single axis.

These results are shown in Figure 4, clearly showing how the behavior is non-linear. The higher infill rate $\rho_{TPU}$, the higher the stresses for the same deformation. These experimental data will allow to fit a non-linear model that will be used to model the mechanical behavior of the arm.

Finally, for   additive   manufacturing,   the   Creality   CR10   S5 printer   has   been selected   due   to   its   large   dimensions (500x500x500 mm), allowing the printing of the arms in any desired orientation, since this affects the arrangement of the fibers. The basic printer requirements for TPU are a extruder temperature of 225-250°C, a heated print bed of 50 ºC, and a direct drive extruder.


\begin{figure}
\includegraphics[width=0.47\textwidth,scale=0.25]{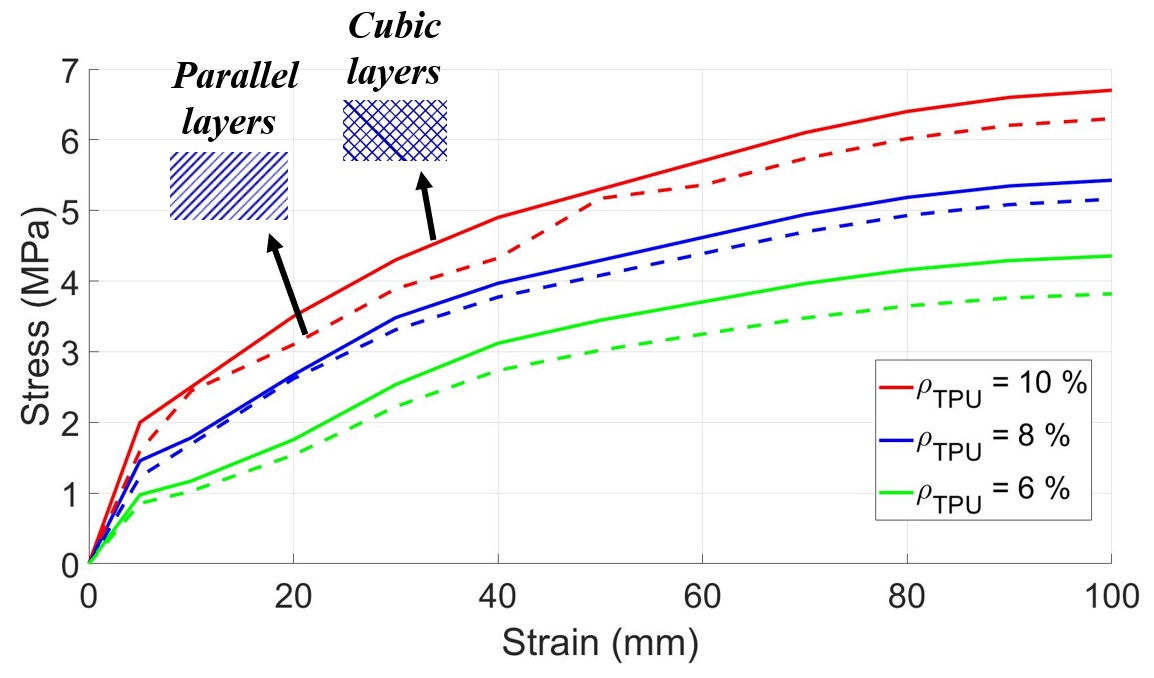}
\caption{TPU 70A stress-strain ($\sigma - \epsilon$) curves for different internal densities $\rho_{TPU}$ and grid patterns}
\label{sigma}
\end{figure}



\subsection{Soft arm design}

For the design of the arm, theoretical analyses, mechanical simulations (structural and computational fluid dynamics), and experiments on a test bench have been carried out. In parallel, research has been carried out on the manufacturing methodology. 


The propulsive equipment is composed of an electric motor (DJI
2312E, providing 450g of nominal thrust using a 4S (14.8V) battery), an electronic speed controller (DJI 430 Lite ESC) and a 10-inch DJI plastic propeller to make it safer than carbon fiber. The arm is also equipped with soft tendons on its underside, responsible for generating compression forces that cause the arm to bend downwards. These tendons are composed of nylon threads that are wound on a 3D-printed reel actuated by a HITEC MG996R servomotor which provides a maximum torque of 35 kgcm. 

Tests have also been carried out to investigate the possibility of replacing the servomotor with Shape Memory Alloys (SMA's). However, the use of SMAs has been discarded for several reasons, not only related to the actuator force, which is smaller in this case. First, power electronics required for its control are excessively heavy and complex today. But most importantly, thermal dynamics of SMA's are slow, which would prevent an immediate opening of the arm in an emergency situation.


Finally, the arm is equipped with an FRS contact force sensor from the manufacturer Interlink Electronics, which allows analyzing the level of attachment to the pipe. On the other hand, an inertial measurement unit (IMU BNo055) has been used to determine the angle of deflection of the arm $\alpha_{a_rm}$.

Regarding geometry, the arm is designed with an initial deflection downwards $\alpha_0$, so that when the motors are turned on, it deflects upwards and remains approximately horizontal at 50$\%$ throttle. Arm deflections in flight might involve control difficulties and thrust losses in a future UAV prototype that are unavoidable if flexible materials are used. Nevertheless, it is possible to dampen them, especially the upward bending. This is achieved thanks to a suitable design, increasing the internal density in the vicinity of the upper surface and the thickness of the upper wall with respect to the lower one. It is also achieved by arranging TPU fibers parallel to the axis of the arm in the upper area and perpendicularly in the lower area.

On the other hand, the lower surface includes curvatures or folds that allow the tendons to bend the arm downwards. The design of the angles and lengths of said curvatures in order to adapt perfectly to pipes of different diameters is a problem studied by the authors in previous articles \cite{GRubiales2021}, and the optimal values are shown in Table I. 

\begin{table}[H]
\centering
\begin{tabular}{c c c c c c c c}
\hline
$\beta_1$ & $L_1$ & $\beta_2$ & $L_2$ & $\beta_3$ & $L_3$ & $\beta_4$ & $L_4$ \\
$36^o$ & $35mm$ & $27^o$ & $40mm$ & $19^o$ & $45mm$ & $13^o$ & $55mm$ \\
\hline
\end{tabular}
\caption{Geometry of the lower surface of the arm: inclination angles of each fold ($\beta_i$) and lengths ($L_i$) }
\end{table}

\section{Mechanical modeling}

\subsection{Structural analysis}

In the design of typical structures used in UAVs, it is common to propose simplified material resistance models to approximately calculate stresses and displacements. However, the flexibility of the arm proposed in this work, together with the complexity of the geometry and the distribution of forces, make it necessary to solve the elastic problem without simplifications.

\begin{figure}
\includegraphics[width=0.47\textwidth,scale=0.25]{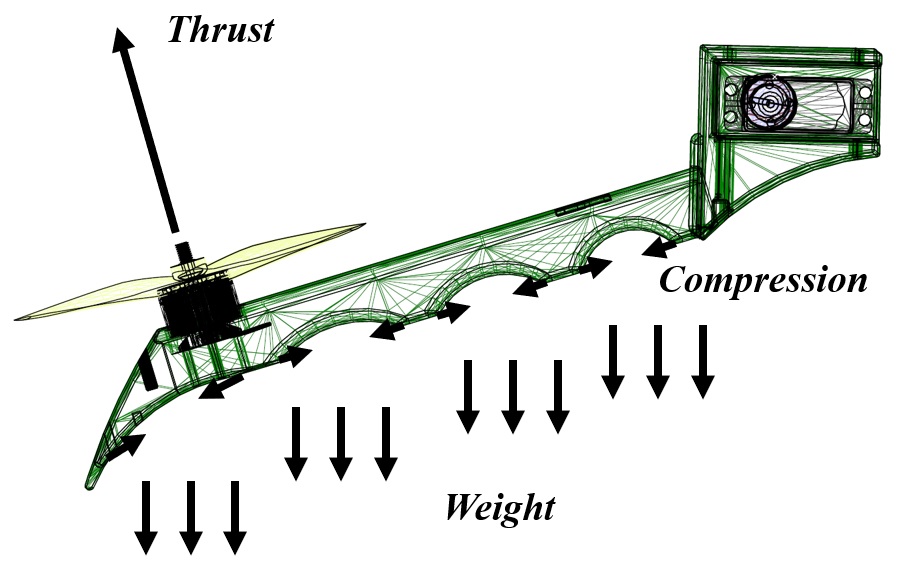}
\caption{Diagram of the forces acting on the soft propelled arm: thrust, weight and tendons compression }
\label{angles}
\end{figure}

The present linear elastic problem is defined by the geometry of the arm, the properties of the material (studied in Section II.B), the external forces acting (which are described in Figure 4: thrust, weight and tendons compression) and boundary conditions (a fixed support at the base of the arm). From these elements it is possible to find an internal stress field (the six components of the stress tensor $\sigma_x, \sigma_y, \sigma_z, \sigma_{xy}, \sigma_{xz}$ and $\sigma_{yz}$), a displacement field ($u_x$, $u_y$ and $u_z$), and also the strain field ($\epsilon_x, \epsilon_y, \epsilon_z, \epsilon_{xy}, \epsilon_{xz}$ and $\epsilon_{yz}$). This information is essential to know the points of the arm that suffer bigger stresses in order to reinforce them (for example, increasing TPU density in that area) as well as to know the deflections of the arm in flight conditions.

These 15 unknowns can be determined from the following equilibrium and compatibility equations, that allow to integrate the external forces $X_i$ in the problem and establish a relation between the deformations and the displacements. However, there are still 6 extra equations that are determined by the constitutive equations of the material.

- Cauchy's equilibrium equations:

\begin{equation}
\sigma_{ij,j}+X_i=0
\end{equation}

- Saint-Venant's compatibility equations:

\begin{equation}
\epsilon_{ij}=\frac{1}{2}(u_i,_j+u_j,_i)
\end{equation}

For a homogeneous isotropic linear elastic material these constitutive equations are given by the Lamé-Hooke equations, where E is the Young's modulus, G is the transverse elasticity modulus and $\nu$ is the Poisson's ratio. 

\begin{equation}
\sigma_{ij}=2G\epsilon_{ij} + \lambda \epsilon_{kk}\delta_{ij}
\end{equation}

\begin{equation}
\lambda=\frac{E\nu}{(1+\nu)(1-2\nu)}
\end{equation}

This model is acceptable for not very low internal densities $\rho_{TPU}$. As the flexibility increases, the behavior becomes strongly non-linear under compressive loads that cause bending, and the deformations become noticeable under moderate thrust values. Abandoning the assumption of small deformations requires the use of a nonlinear deformation tensor, specifically it is a quadratic application of the deformation gradient

\begin{equation}
E=\frac{1}{2}(\nabla u^T +\nabla u + \nabla u^T\nabla u)
\end{equation}

Finally, the constitutive equations of the material must also be adapted to the non-linear case. Isotropy hypothesis should also be eliminated, since, except in the case of internal cubic pattern, the fibers during 3D printing are not isotropically arranged, as can be seen in Figure 2. Therefore, the most appropriate approach to close the problem is in energetic terms, through a strain energy density function W.

\begin{equation}
W= \sum_{p,q=0}^{N=2}C_{pq}(I_1-3)^p(I_2-3)^q
\end{equation}

This methodology is known as the 5-parameter Mooney-Rivlin model, and is commonly used to model the elastic response of rubber-like materials. For the resolution of these equations through said model, the finite element method has been used through the Ansys Mechanics commercial software.

This requires knowledge of these Mooney-Rivlin parameters for the material considered in this work. Said parameters are obtained by polynomial adjustment from the non-linear stress-strain curves of Figure 4. The results obtained from said adjustment for different internal densities of the material are shown in Table II.

\begin{table}[H]
\centering
\begin{tabular}{c c c c c c}
\hline
$\rho_{TPU} (\%)$ & $C10$ & $C01$ & $C20$ & $C02$ & $C11$ \\
\hline
6 & -3.19 & 4.23 & 0.64 &  -2.65 & 4.37\\
8 & -4.07 & 4.18 & 0.71 &  -2.62 & 4.54 \\
10 & -4.51 & 4.16 & 0.76 &  -2.75 & 4.89 \\
\hline
\end{tabular}
\caption{Experimentally-measured Mooney-Rivlin coefficients for TPU 70A and different internal densities $\rho_{TPU}$}
\end{table}

Once this model is defined, the aforementioned elastic problem can be solved. In this work, static structural analyzes of the arm are carried out, for different values of the forces, and for different types of fiber distribution in the arm (that is, the arm has been modeled internally in CAD, not only as solid).

\begin{figure}
\includegraphics[width=0.47\textwidth,scale=0.25]{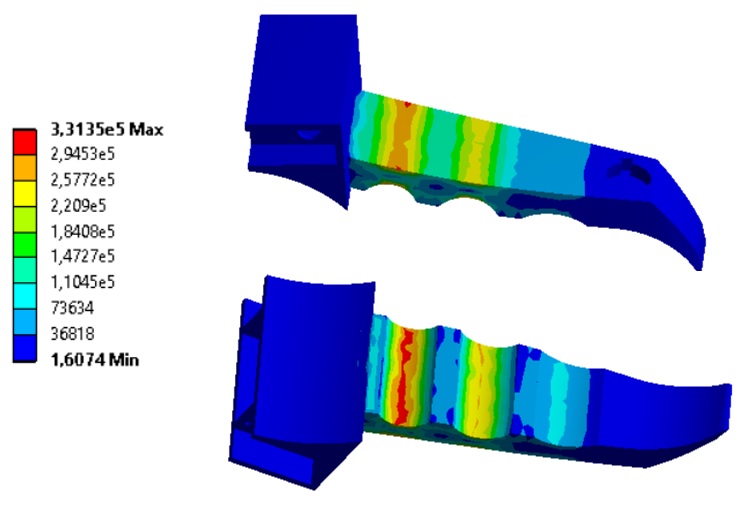}
\caption{Stress distribution (in Pascals) in the arm when a thrust force of 3N is applied for an internal cubic pattern and $\rho_{TPU} = 10 \%$.}
\label{ansys_1}
\end{figure}

Regarding the upward bending of the arm (Figure \ref{ansys_1}), it is observed that the maximum stresses are found in the area of the first fold, and it is therefore the area with the greatest risk. Therefore, the design is adapted by locally increasing the internal density of the material.

\begin{figure}
\includegraphics[width=0.47\textwidth,scale=0.25]{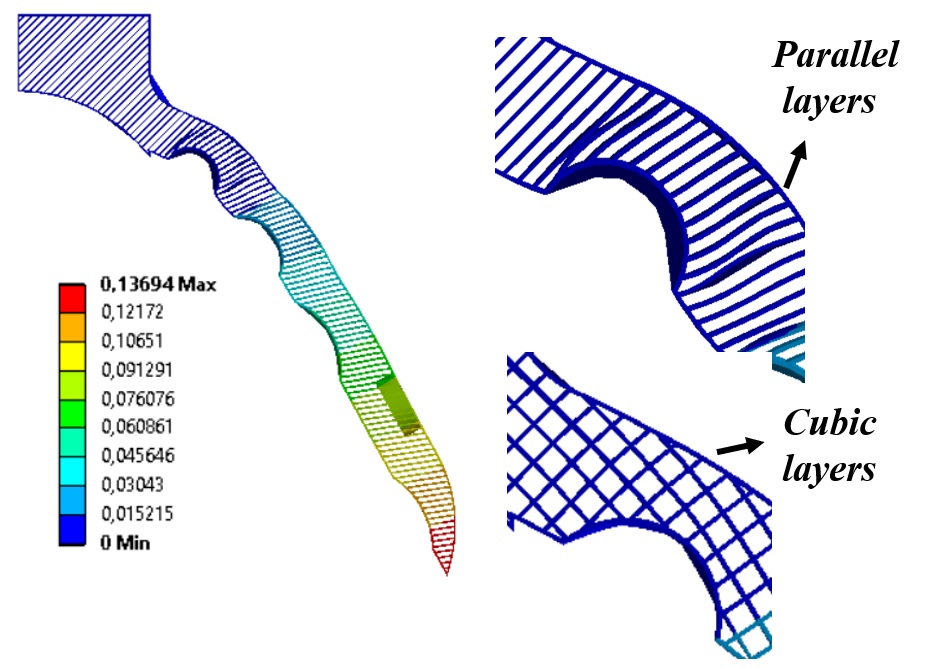}
\caption{Maximum strains (in meters) for a downward deflection of the arm as a reaction to the first tendon compression force. Comparison of the internal structure bending for parallel and cubic layers for $\rho_{TPU} = 7.5 \%$.}
\label{ansys_2}
\end{figure}

In Figure \ref{ansys_2}, the downward bending of the arm is modeled, specifically the action of the first tendon. In this case the deformations become non-linear. The arrangement of the fibers greatly influences this mechanism. The zig-zag (parallel layers) pattern causes a higher local stress concentration than the cubic case, which results in a more distributed stress distribution. However, the zig-zag arrangement causes a greater deflection for the same force, which is favorable since it reduces the energy consumption of the servomotor. The proposed solution consists of applying the zig-zag grid and reinforcing said area with a higher density.

\begin{figure}
\includegraphics[width=0.47\textwidth,scale=0.25]{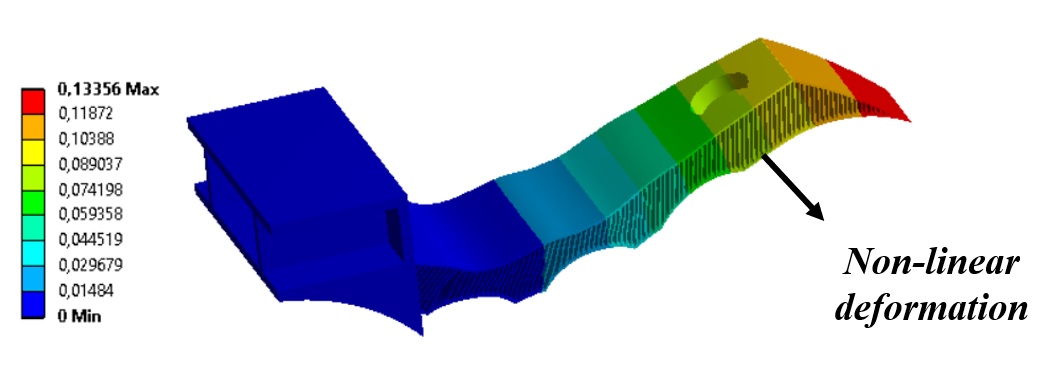}
\caption{Maximum strains (in meters) for a non-linear upward bending of the arm for a small infill pattern $\rho_{TPU}= 5 \%$, and a thrust force of 7 N (around 80 \% throttle).}
\label{ansys_3}
\end{figure}

Finally, these non-linear effects can also appear in upward bending, as shown in Figure \ref{ansys_3}. This occurs for the most flexible cases, with $\rho_{TPU} < 5\%$, and at high thrust values. This area must be avoided since it will cause control difficulties for the UAV in flight.

\subsection{CFD analysis}

This section aims to analyze the efficiency of the proposed design (considering thrust loss due to the arm interaction) and other aerodynamic interactions of the flow. These complex fluid analyses are tipically performed using Computational Fluid Dynamics (CFD), which allows the modelling of the turbulent flow around a propeller, and the interaction of said flow with the arm. 

In this work, the numerical simulations have been performed using the
commercial software Ansys Fluent.
The governing equations of fluid mechanics are based on the conservation of mass, momentum and energy. These equations can be simplified since the flow around a propeller is incompressible ($M<0.3$). Under this consideration, the
Navier-Stokes formulation of the fluid problem is:

\begin{equation}
\nabla \vec{v}= 0
\end{equation}

\begin{equation}
\rho \frac{\partial \vec{v}}{\partial t}+\rho (\vec v \nabla \vec v)= -\nabla p + \mu \nabla^2 \vec v + \rho g
\end{equation}

Being a mechanical analysis, the energy equation is not considered. In these equations, $p$ is the static pressure, $\rho$ is the density, $\vec v$ is the
flow velocity, $\rho g$ are the gravitational forces, and $\mu$ is the viscosity. Each of these variables (3 velocity components and pressure) are decomposed into mean and
fluctuating parts, and the  time-averaged equations are solved.

\begin{equation}
v= \hat{v}+v'
\end{equation}

In these new equations, a new term, known as Reynolds stress tensor ($-\rho \bar{v_i'} \bar{v_j'}$) plays a fundamental role. This new set of equations is called Reynolds-Averaged Navier Stokes (RANS)
equations. The system is closed with a $k-\epsilon$ turbulent model with standard wall
functions. This model is especially suitable for flows involving rotation and recirculations. 

\begin{figure}
\includegraphics[width=0.47\textwidth,scale=0.25]{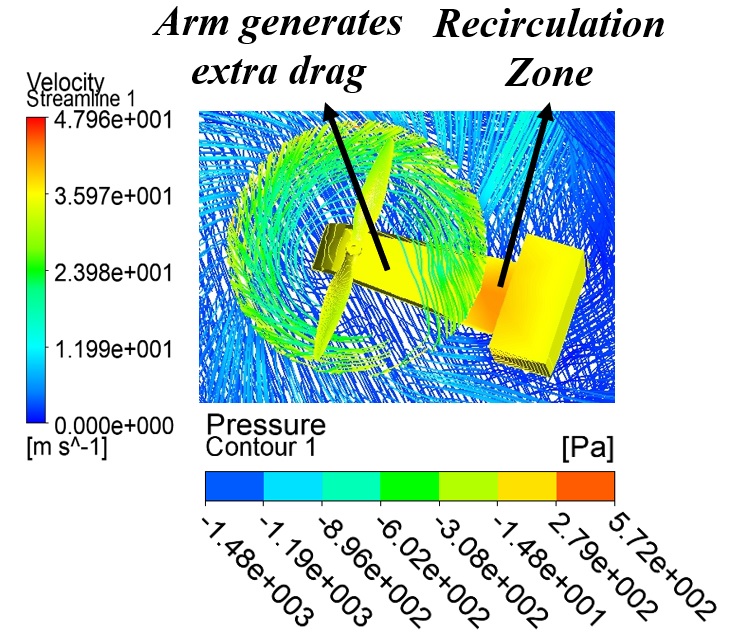}
\caption{CFD analysis of the soft propelled arm at $T \approx 50\%$ throttle, with $\alpha_{arm} \approx 0^o$. Velocity streamlines for both the static and rotational domains are provided together with the pressure distributions in the arm.}
\label{angles}
\end{figure}

The behavior of the propelled arm is studied through a rotational domain, whose rotational speed corresponds to the engine thrust. The rest of the fluid domain, in which the arm is located, is static. The goal is to carry out a comparison between the thrust provided by an isolated propeller, and the thrust of the propeller integrated in the arm (see Eq. 12). This analysis is repeated for various designs depending on the position of the motor in the arm (Figure 12) and for various angles of deflection of the arm (Figure 13). In turn, it is performed for several different rotational speeds.

\begin{equation}
\eta=\frac{F_P}{F_{P+A}}
\end{equation}

Figures 9, 10 and 11 show the velocity streamlines and pressure distributions in the arm for the three configurations studied. First, in the CFD analysis of the soft propelled arm at $T \approx 50\%$ throttle, with $\alpha_{arm} \approx 0^o$ (Figure 9), the main propeller-arm interaction is aerodynamic resistance, with recirculation being practically negligible. However, as the arm deflects upwards (Figure 10), the recirculation increases and the so-called corner-effect becomes more evident. Therefore, for the same absolute deflection, the efficiency is greater if the arm is deflected upwards rather than downwards.

For $\alpha_{arm} \neq 0^o$ the main efficiency loss is due to the fact that the thrust is not only directed along the z axis, but is also projected along the y axis (Figure 13). For this reason, it is essential to focus the design on limiting and damping the deflections of the arm in the operating range of the motor. This will also have implications on the dynamics of a prototype UAV in which these arms are integrated, as it alters the controller's mixer matrix with the deflections of the arms in flight.

\begin{figure}
\includegraphics[width=0.47\textwidth,scale=0.25]{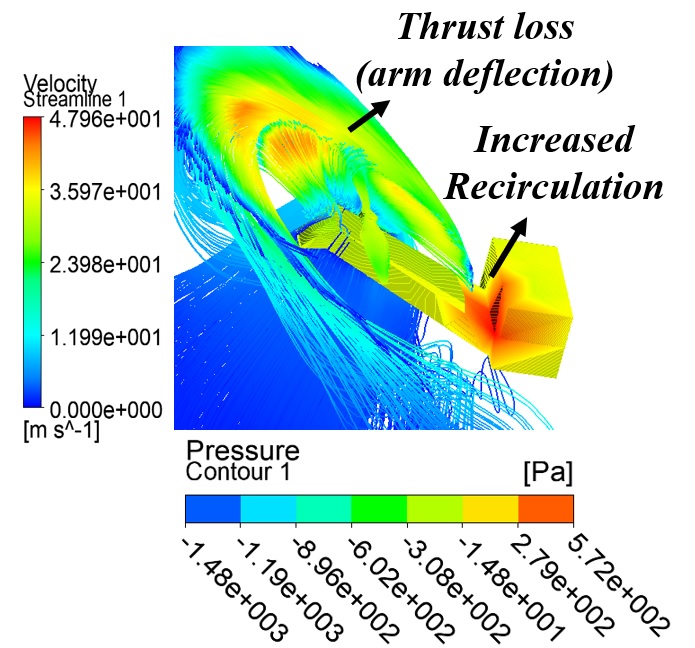}
\caption{CFD analysis of the soft propelled arm at $T \approx 100\%$ throttle, with $\alpha_{arm} \approx 30^o$. Velocity streamlines for both the static and rotational domains are provided together with the pressure distributions in the arm.}
\label{angles}
\end{figure}

\begin{figure}
\includegraphics[width=0.47\textwidth,scale=0.25]{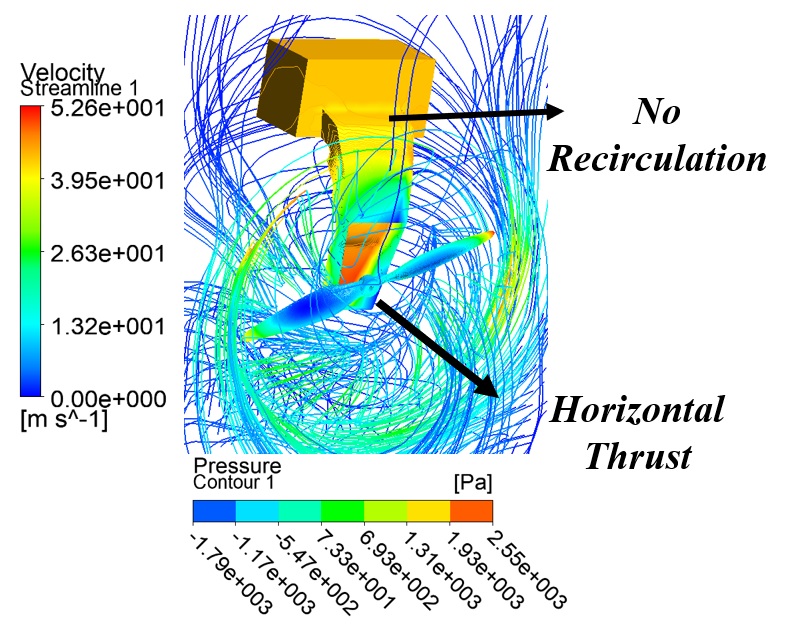}
\caption{CFD analysis of the soft propelled arm in the take-off phase after attachment to the pipe. Velocity streamlines for both the static and rotational domains are provided together with the pressure distributions in the arm.}
\label{angles}
\end{figure}

Figure 11 shows a CFD analysis of the soft propelled arm in the take-off phase after attachment to the pipe. In this case, all the thrust is directed along the y axis, and there is no recirculation.

Finally, Figure 12 shows that, as the propeller gets closer to the base of the arm, the efficiency decreases as aerodynamic drag is greater (the width of the arm increases). However, the existing recirculation in the vertical surface of the arm base generates an extra thrust, which stops being significant from $x/c \approx 0.8$. These opposing effects give rise to the existence of an optimum $\frac{x}{c}*$, which moves away from the base as the rotational speed is greater, since the recirculation is dependent on the level of turbulence. For the design, the optimum corresponding to the nominal thrust of the engines ($\approx 500g$, around $4000 rpm$), which is $x/c*=0.83$, has been chosen.

Table III shows a comparison of the thrust efficiencies obtained for the optimal configuration simulated above, for inclinations between $\alpha_{arm}=\pm20 ^o$, depending on the rotational speed. It can therefore be concluded that the efficiencies are greater than 90 $\%$ for the main operating points of the propeller.

\begin{table}[H]
\centering
\begin{tabular}{c c }
\hline
Rotational speed (rpm) & $\eta$ \\
\hline
6000 & 91.6 \% \\
5000 & 90.9 \%  \\
4000 & 89.5 \% \\
\hline
\end{tabular}
\caption{Thrust efficiencies for the optimum configuration $(x/c*=0.83)$ and ($\alpha_{arm}<20^0$) for different rotational velocities (rpm)}
\end{table}

\begin{figure}
\includegraphics[width=0.47\textwidth,scale=0.25]{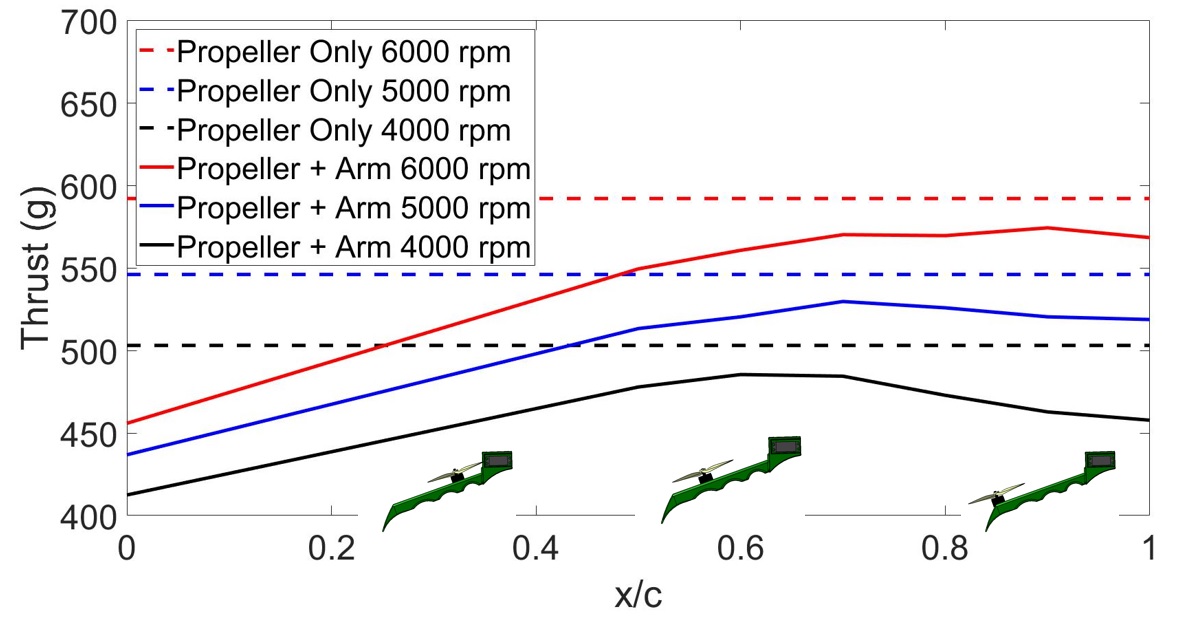}
\caption{Evolution of the simulated thrust force for an isolated propeller (dashed lines) and a propeller interacting with the arm (solid lines) for different rotational velocities as a function of the position of the motor x/c}
\label{angles}
\end{figure}

\begin{figure}
\includegraphics[width=0.47\textwidth,scale=0.25]{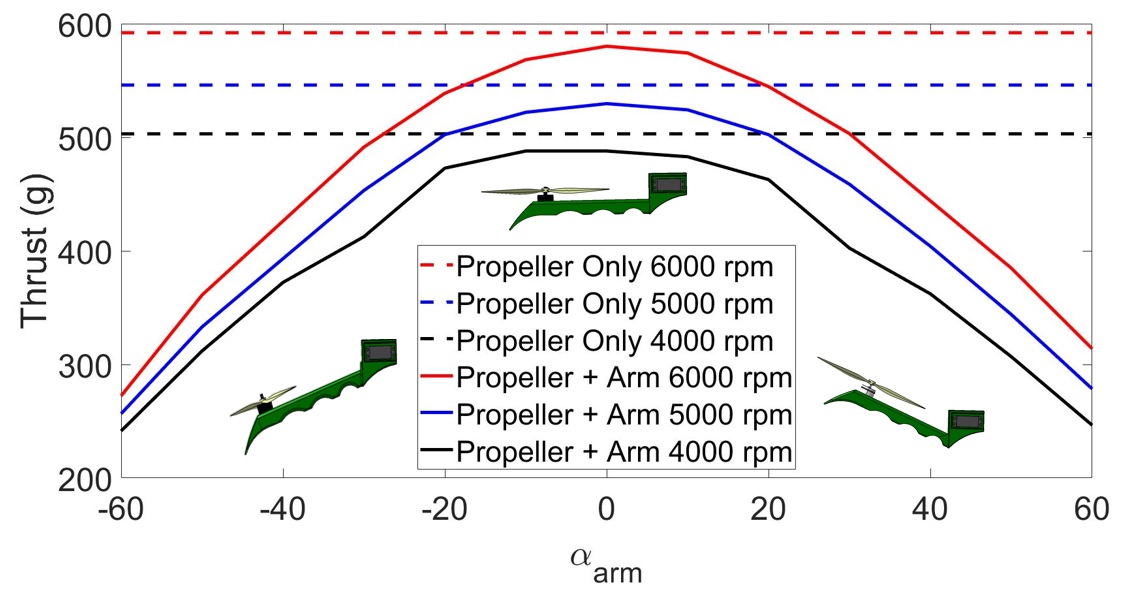}
\caption{Evolution of the simulated thrust force for an isolated propeller (dashed lines) and a propeller interacting with the arm (solid lines) for different rotational velocities as a function of the deflection of the arm $\alpha_{arm}$}
\label{angles}
\end{figure}

\section{Results}

\subsection{Experimental setup}

The soft propelled arm has been fixed to a pipe at its base with the aim of studying the behavior in conditions similar to its integration in a UAV (Figure 14). Experiments have been carried out varying the rotational speeds of the propellers, characterizing the deflections of the arm through the Inertial Measurements Unit (IMU). Pipe-attachment experiments have also been carried out, analyzing the grip force through a contact force FRS sensor. These experiments are the previous step to the integration of the soft propelled arm in a UAV. In the next section, a comparison of the results obtained through numerical simulations with the experimental results is performed.

\begin{figure}[!tbp]
 \centering
    \begin{subfigure}[Natural state]{
     \begin{minipage}[t]{0.43\linewidth}
     \centering
     \includegraphics[width=0.9\textwidth]{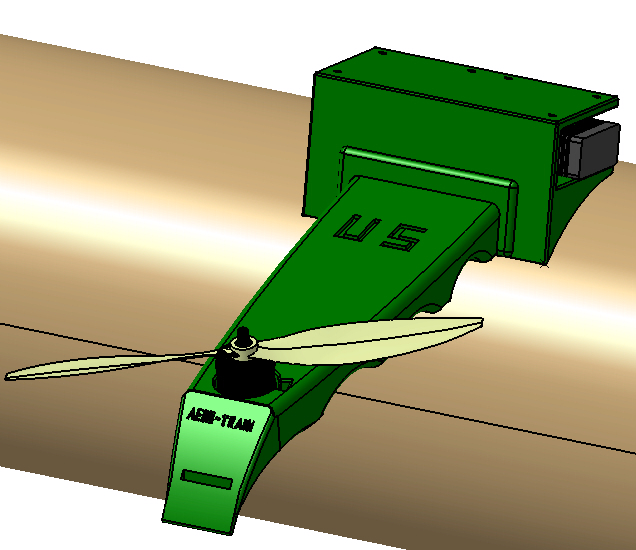}
     \end{minipage}%
     }%
    \end{subfigure}
    \begin{subfigure}[Attached to the pipe]{
     \begin{minipage}[t]{0.43\linewidth}
     \centering
     \includegraphics[width=0.7\textwidth]{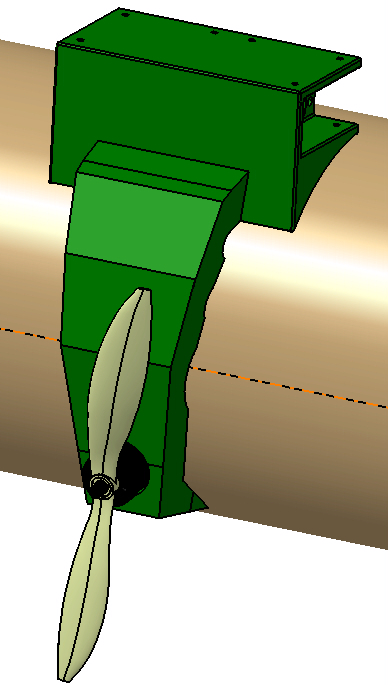}
     \end{minipage}%
     }%
    \end{subfigure}
    \begin{subfigure}[At 100\% throttle]{
     \begin{minipage}[t]{0.96\linewidth}
     \centering
     \includegraphics[width=0.8\textwidth]{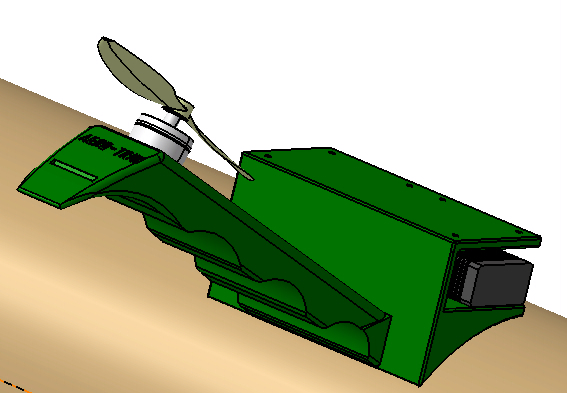}
     \end{minipage}%
     }%
    \end{subfigure}
 \caption{CAD model of the soft arm in the experimental test bench}
 \label{f:arm}
\end{figure}

\subsection{Results comparison}

In Figure 15, it is observed that the deflections of the arm upwards are smaller than downwards (the slope of the curve for $\alpha_{arm}>0$ is smaller). This is achieved thanks to the aforementioned suitable design of the arm.

For small internal densities $\rho_{TPU}$, and high throttle values, the arm changes from a linear elastic behavior to being strongly non-linear. For instance, in the case $\rho_{TPU} = 5\%$, it is unfeasible to work at throttles greater than 80$\%$, since this zone is extremely complicated to model, with large deflections for small increases in force. For the almost linear-elastic region, the following model is proposed, which constants A1, A2, B1 and B2 are determined by polynomial interpolation from Figure 15. The interpolated values are shown in Table IV.

\begin{equation}
\alpha_{arm} = \alpha(0)+(A1+\rho_{TPU}*A2)*T+(B1+\rho_{TPU}*B2)T^2
\end{equation}

\begin{table}[H]
\centering
\begin{tabular}{c c c c}
\hline
A1 & A2 & B1 & B2 \\
2.4387 & -0.1997 & -0.162 & 0.0151 \\
\hline
\end{tabular}
\caption{Arm deflections model coefficients for Equation 13}
\end{table}

The numerical and experimental results are practically similar (within experimental uncertainties) especially in the linear zone. In the non-linear zone, the differences become more evident. This is due to the complexity of modeling the behavior of this type of material, even after using a model as advanced as the Mooney-Rivlin.

\begin{figure}
\includegraphics[width=0.47\textwidth,scale=0.25]{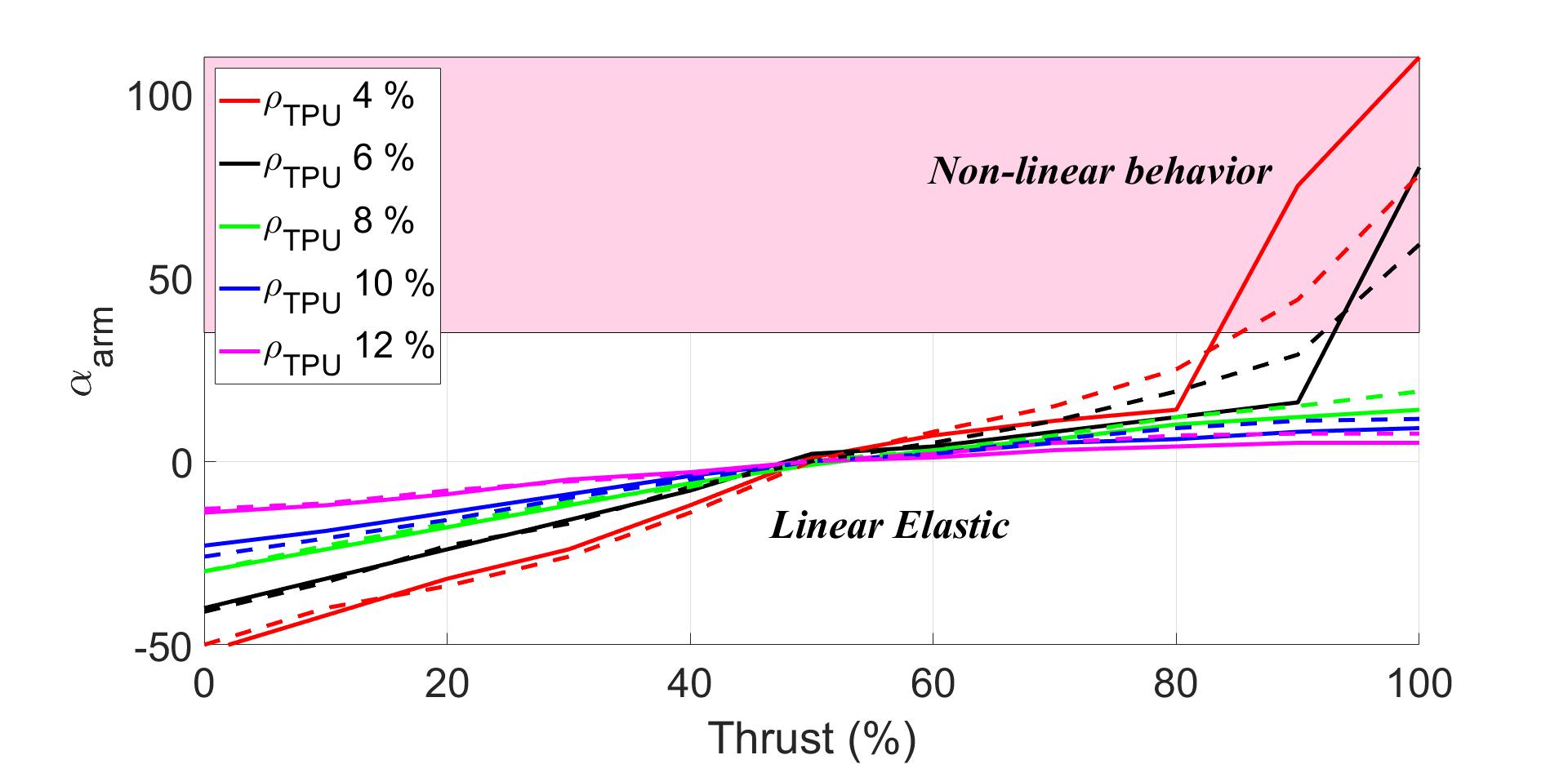}
\caption{Experimentally measured (solid lines) and simulated (dashed lines) arm deflections $\alpha_{arm}$ as a function of thrust (\%) for different internal densities $\rho_{TPU}$. }
\label{angles}
\end{figure}

Finally, the contact force with the pipe is analyzed for different values of the infill rate $\rho_{TPU}$. It can be seen that for infill rates lower than $\rho_{TPU}<15\%$ (this limit roughly coincides with the numerical results) complete bending is already produced, although a minimum force of about 1000 N/m2 is necessary to guarantee correct adherence.

\begin{figure}
\includegraphics[width=0.47\textwidth,scale=0.25]{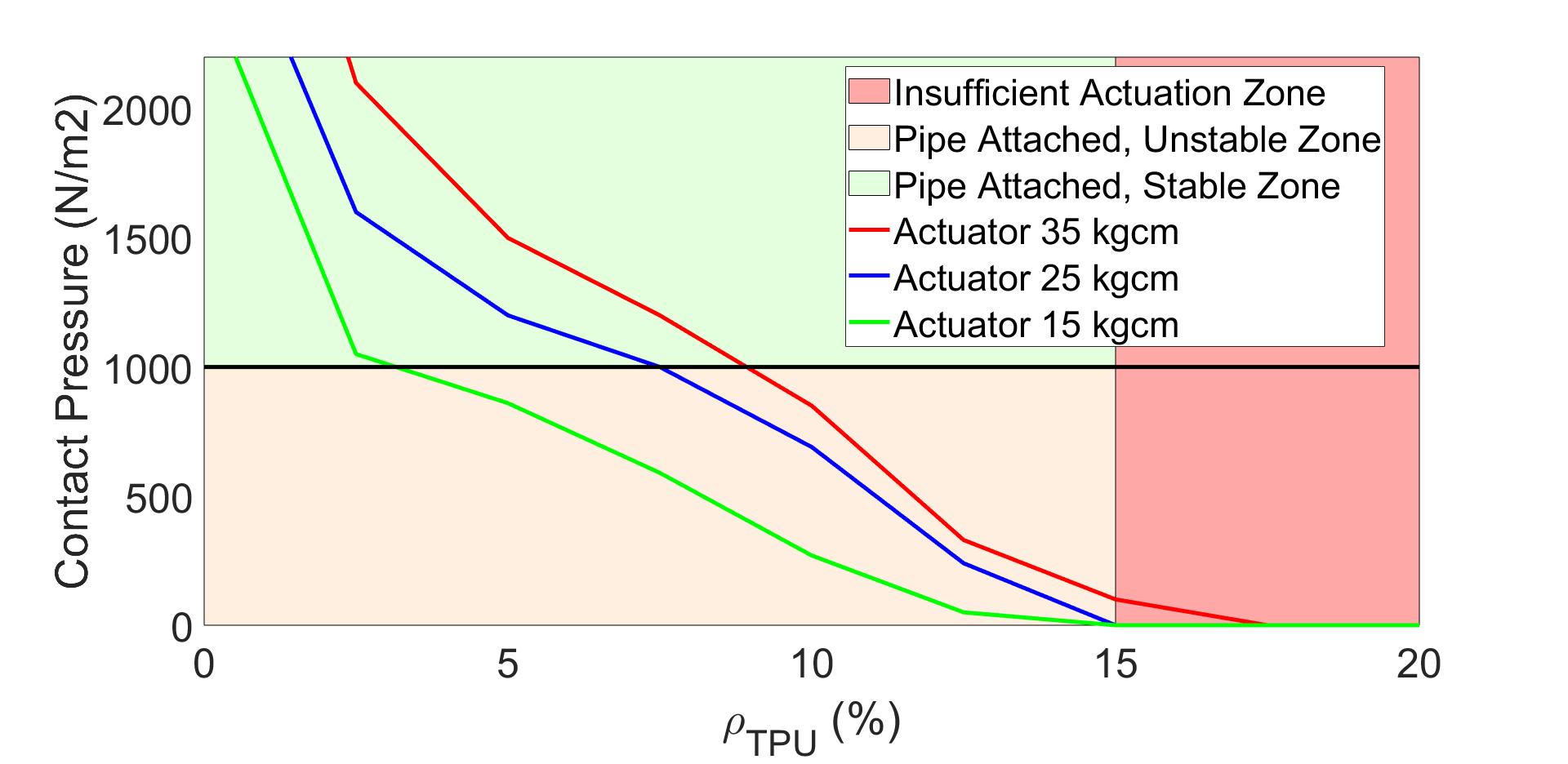}
\caption{Contact pressure forces (N/m2) for different actuator forces as a function of infill rate $\rho_{TPU}$. In the red area, the arm is too rigid to bend and attach to the pipe.}
\label{angles}
\end{figure}

In this way, it can be concluded that the optimum density range for $\rho_{TPU}$ is between 6 and 8 $\%$, since the arm is sufficiently flexible to adapt to the environment (pipes), while the deflections are sufficiently small and with an easy-to-model, almost linear behavior.

\section{CONCLUSIONS}

In this paper, the use of flexible materials in UAVs is proposed. Specifically, a flexible propelled arm, which has the ability to deform and adapt to pipes of different diameters,  is presented.

The design has fulfilled the objectives pursued. Thus,  it minimizes upward deflection of the arm  and the arm deflection is less than 14 degrees for $\rho_{TPU}>6\%$ in the operating range of the motors. This is essential to ensure the flyability of the future UAV prototype. In addition, thrust efficiencies are above 90 \% in the propeller operation range. In certain arm configurations, the interaction with the body of the arm can be used to generate extra thrust.

The path of the authors research goes towards the implementation of these developments in a complete UAV, within the range of densities aforementioned, while guaranteeing the flyability of the vehicle through the appropriate control techniques. In the long term, the goal is to develop machine learning techniques that allow for an even greater flexibility. A data base approach, better than a model based approach,  will be used.

This design is just a first step towards soft UAVs, which will impact the acceptance of UAVs  collaborating with humans in daily tasks.

\section*{ACKNOWLEDGMENTS}

This work has been developed within the framework of the AERO-TRAIN project (Aerial Robotic Training for the next generation of European infrastructure and asset maintenance technologies), a Marie-Sklodowska-Curie Innovative Training Network (ITN) - European Training Network (ETN) project  (Grant agreement 953454). The  funding  of the  HYFLIERS  project  (H2020-2017-779411)  is  also  acknowledged. We thank Robotics, Vision and Control Group (GRVC), with special thanks to Abraham López Lora, for supporting us during this work.


\bibliographystyle{IEEEtran}
\bibliography{References}

\begin{thebibliography}{10}
\providecommand{\url}[1]{#1}
\csname url@rmstyle\endcsname
\providecommand{\newblock}{\relax}
\providecommand{\bibinfo}[2]{#2}
\providecommand\BIBentrySTDinterwordspacing{\spaceskip=0pt\relax}
\providecommand\BIBentryALTinterwordstretchfactor{4}
\providecommand\BIBentryALTinterwordspacing{\spaceskip=\fontdimen2\font plus
\BIBentryALTinterwordstretchfactor\fontdimen3\font minus
  \fontdimen4\font\relax}
\providecommand\BIBforeignlanguage[2]{{%
\expandafter\ifx\csname l@#1\endcsname\relax
\typeout{** WARNING: IEEEtran.bst: No hyphenation pattern has been}%
\typeout{** loaded for the language `#1'. Using the pattern for}%
\typeout{** the default language instead.}%
\else
\language=\csname l@#1\endcsname
\fi
#2}}

\bibitem{Bernard2011AutonomousTA}
M.~Bernard, K.~Kondak, I.~Maza, and A.~Ollero, ``Autonomous transportation and
  deployment with aerial robots for search and rescue missions,'' \emph{J.
  Field Robotics}, vol.~28, pp. 914--931, 2011.

\bibitem{Ollero2012}
L.~Merino, F.~Caballero, J.~R. Martinez-de Dios, I.~Maza, and A.~Ollero, ``An
  unmanned aircraft system for automatic forest fire monitoring and
  measurement,'' \emph{Journal of Intelligent and Robotic Systems}, vol.~65,
  pp. 533--548, 01 2012.

\bibitem{OlleroManipulation}
F.~Ruggiero, V.~Lippiello, and A.~Ollero, ``Aerial manipulation: A literature
  review,'' \emph{IEEE Robotics and Automation Letters}, vol.~3, pp.
  1957--1964, 07 2018.

\bibitem{Rao2016}
B.~Rao, A.~G. Gopi, and R.~Maione, ``The societal impact of commercial
  drones,'' \emph{Technology in Society}, vol.~45, pp. 83--90, 05 2016.

\bibitem{aerialmanipulator}
G.~Giglio, F.~Pierri, M.~Trujillo, G.~Antonelli, F.~Caccavale, A.~Viguria,
  S.~Chiaverini, and A.~Ollero, ``Behavioral control of unmanned aerial vehicle
  manipulator systems,'' \emph{Autonomous Robots}, vol.~41, pp. 1203--1220,
  2017.

\bibitem{Zheng2021EvolutionaryHC}
Y.~Zheng, Y.-C. Du, Z.-L. Su, H.~Ling, M.-X. Zhang, and S.~Chen, ``Evolutionary
  human-uav cooperation for transmission network restoration,'' \emph{IEEE
  Transactions on Industrial Informatics}, vol.~17, pp. 1648--1657, 2021.

\bibitem{ZHAO2020}
\BIBentryALTinterwordspacing
Z.~ZHAO, Y.~NIU, and L.~SHEN, ``Adaptive level of autonomy for human-uavs
  collaborative surveillance using situated fuzzy cognitive maps,''
  \emph{Chinese Journal of Aeronautics}, vol.~33, no.~11, pp. 2835--2850, 2020,
  sI: Emerging Technologies of Unmanned Aerial Vehicles. [Online]. Available:
  \url{https://www.sciencedirect.com/science/article/pii/S1000936120302181}
\BIBentrySTDinterwordspacing

\bibitem{Trivedi2008SoftRB}
D.~Trivedi, C.~D. Rahn, W.~M. Kier, and I.~D. Walker, ``Soft robotics:
  Biological inspiration, state of the art, and future research,''
  \emph{Applied Bionics and Biomechanics}, vol.~5, pp. 99--117, 2008.

\bibitem{Rus2015DesignFA}
D.~Rus and M.~T. Tolley, ``Design, fabrication and control of soft robots,''
  \emph{Nature}, vol. 521, pp. 467--475, 2015.

\bibitem{Riviere2018AgileRF}
V.~Riviere, A.~Manecy, and S.~Viollet, ``Agile robotic fliers: A morphing-based
  approach,'' \emph{Soft Robotics}, vol.~5, pp. 541 -- 553, 2018.

\bibitem{Ajanic2020BioInspiredSW}
E.~Ajanic, M.~Feroskhan, S.~Mintchev, F.~Noca, and D.~Floreano, ``Bio-inspired
  synergistic wing and tail morphing extends flight capabilities of drones,''
  \emph{arXiv: Fluid Dynamics}, 2020.

\bibitem{ReversiblePlasticity}
\BIBentryALTinterwordspacing
D.~Hwang, E.~J. Barron, A.~B. M.~T. Haque, and M.~D. Bartlett, ``Shape morphing
  mechanical metamaterials through reversible plasticity,'' \emph{Science
  Robotics}, vol.~7, no.~63, p. eabg2171, 2022. [Online]. Available:
  \url{https://www.science.org/doi/abs/10.1126/scirobotics.abg2171}
\BIBentrySTDinterwordspacing

\bibitem{Floreano_origami}
\BIBentryALTinterwordspacing
S.~Mintchev, J.~Shintake, and D.~Floreano, ``Bioinspired dual-stiffness
  origami,'' \emph{Science Robotics}, vol.~3, no.~20, p. eaau0275, 2018.
  [Online]. Available:
  \url{https://www.science.org/doi/abs/10.1126/scirobotics.aau0275}
\BIBentrySTDinterwordspacing

\bibitem{iros-2016}
\BIBentryALTinterwordspacing
\emph{2016 IEEE/RSJ International Conference on Intelligent Robots and Systems,
  IROS 2016, Daejeon, South Korea, October 9-14, 2016}.\hskip 1em plus 0.5em
  minus 0.4em\relax IEEE, 2016. [Online]. Available:
  \url{http://ieeexplore.ieee.org/xpl/mostRecentIssue.jsp?punumber=7743711}
\BIBentrySTDinterwordspacing

\bibitem{Xiang2019SoftsmartRE}
C.~Xiang, J.~Guo, and J.~M. Rossiter, ``Soft-smart robotic end effectors with
  sensing, actuation, and gripping capabilities,'' \emph{Smart Materials and
  Structures}, 2019.

\bibitem{TammTCP}
A.~E. Gomez-Tamm, P.~Ramon-Soria, B.~C. Arrue, and A.~Ollero, ``Tcp muscle
  tensors: Theoretical analysis and potential applications in aerial robotic
  systems,'' \emph{Iberian Robotics conference}, 2019.

\bibitem{GRubiales2021}
\BIBentryALTinterwordspacing
F.~J. Garcia~Rubiales, P.~Ramon~Soria, B.~C. Arrue, and A.~Ollero,
  ``Soft-tentacle gripper for pipe crawling to inspect industrial facilities
  using uavs,'' \emph{Sensors}, vol.~21, no.~12, 2021. [Online]. Available:
  \url{https://www.mdpi.com/1424-8220/21/12/4142}
\BIBentrySTDinterwordspacing

\bibitem{Mazzolai2012}
B.~Mazzolai, L.~Margheri, M.~Cianchetti, P.~Dario, and C.~Laschi,
  ``Soft-robotic arm inspired by the octopus: Ii. from artificial requirements
  to innovative technological solutions,'' \emph{Bioinspiration \&
  biomimetics}, vol.~7, p. 025005, 06 2012.

\bibitem{TPUproperties}
H.~Lee, R.~i~Eom, and Y.~Lee, ``Evaluation of the mechanical properties of
  porous thermoplastic polyurethane obtained by 3d printing for protective
  gear,'' \emph{Advances in Materials Science and Engineering}, 2019.

\bibitem{lee2017fundamentals}
J.-Y. Lee, J.~An, and C.~K. Chua, ``Fundamentals and applications of 3d
  printing for novel materials,'' \emph{Applied materials today}, vol.~7, pp.
  120--133, 2017.

\bibitem{hyperelastic2}
\BIBentryALTinterwordspacing
A.-A. Said, D.-p. Carlos, and V.~Manuel~F., ``Mechanical assessment and
  hyperelastic modeling of polyurethanes for the early stages of vascular graft
  design,'' \emph{Materials}, vol.~13, no.~21, 2020. [Online]. Available:
  \url{https://www.mdpi.com/1996-1944/13/21/4973}
\BIBentrySTDinterwordspacing

\bibitem{hyperelastic}
\BIBentryALTinterwordspacing
C.~Emminger, U.~D. Çakmak, R.~Preuer, I.~Graz, and Z.~Major, ``Hyperelastic
  material parameter determination and numerical study of tpu and pdms
  dampers,'' \emph{Materials}, vol.~14, no.~24, 2021. [Online]. Available:
  \url{https://www.mdpi.com/1996-1944/14/24/7639}
\BIBentrySTDinterwordspacing

\bibitem{CFD2}
C.~Paz, E.~Suárez, C.~Gil, and C.~Baker, ``Cfd analysis of the aerodynamic
  effects on the stability of the flight of a quadcopter uav in the proximity
  of walls and ground,'' \emph{Journal of Wind Engineering and Industrial
  Aerodynamics}, vol. 206, p. 104378, 11 2020.

\bibitem{CFDmethod}
E.~Loureiro, N.~Oliveira, P.~Hallak, F.~Bastos, L.~Rocha, R.~Delmonte, and
  A.~Lemonge, ``Evaluation of low fidelity and cfd methods for the aerodynamic
  performance of a small propeller,'' \emph{Aerospace Science and Technology},
  vol. 108, p. 106402, 01 2021.

\bibitem{PRamon2019}
P.~Ramon-Soria, A.~Gomez-Tamm, F.~Garcia-Rubiales, B.~Arrue, and A.~Ollero,
  ``Autonomous landing on pipes using soft gripper for inspection and
  maintenance in outdoor environments,'' in \emph{2019 IEEE/RSJ International
  Conference on Intelligent Robots and Systems (IROS)}, 2019, pp. 5832--5839.

\end{thebibliography}

\end{document}